\documentclass[prl,reprint,showpacs]{revtex4-1}
\usepackage{subfigure}
\usepackage[]{graphicx}
\usepackage{textcomp} 
\usepackage{isotope}
\usepackage{color}
\usepackage{hyperref}
\usepackage{svn-multi}
\renewcommand{\textbullet}{$\vcenter{\hbox{\Large$\bullet$}}$}
\usepackage{amssymb}
\usepackage{amsmath} 
\definecolor{darkgreen}{RGB}{0,128,0}

\begin{document}
\title{Dynamics of single neutral impurity atoms immersed in an ultracold gas}

\author{Nicolas Spethmann,$^{1,2}$ Farina Kindermann,$^{1,2}$
Shincy John,$^{1}$ Claudia Weber,$^{1}$ Dieter Meschede$^1$ and Artur Widera$^2$}

\affiliation{$^1$ Institut f\"ur Angewandte Physik, Universit\"at Bonn,
Wegelerstra\ss e~8, 53115 Bonn, Germany}
\affiliation{$^2$ Fachbereich Physik und Landesforschungszentrum OPTIMAS, Universit\"at Kaiserslautern, 67663 Kaiserslautern, Germany}

\begin{abstract}
We report on controlled doping of an ultracold Rb gas with single neutral Cs impurity atoms. Elastic two-body collisions lead to a rapid thermalization of the impurity inside the Rb gas, representing the first realization of an ultracold gas doped with a precisely known number of impurity atoms interacting via $s$-wave collisions. Inelastic interactions are restricted to a single three-body recombination channel in a highly controlled and pure setting, which allows to determine the Rb-Rb-Cs three-body loss rate with unprecedented precision. Our results pave the way for a coherently interacting hybrid system of individually controllable impurities in a quantum many-body system.

\end{abstract}
\pacs{} 
\maketitle

Recent developments in experimental quantum gas research focus on single particle control in a many-body system for detection and engineering of strongly correlated quantum states, including high-resolution imaging of strongly correlated bosonic systems in optical lattices \cite{gericke_high-resolution_2008, bakr_quantum_2009,sherson_single-atom-resolved_2010} or the preparation of a mesoscopic number of degenerate Fermions \cite{serwane_deterministic_2011}.
While impurities have been studied in balanced or imbalanced mixtures \cite{schirotzek_observation_2009, palzer_quantum_2009}, extending this approach to single or few, individually controllable impurities in a quantum gas grants access to a huge number of novel applications proposed. 
In the direction of quantum information processing, atomtronics applications are envisioned with single atoms acting as switches for a macroscopic system in an atomtronics circuit \cite{micheli_single_2004}; two impurity atoms immersed in a quantum gas can entangle by an effective long-range interaction mediated by the gas \cite{klein_interaction_2005}, or individual qubits can be cooled preserving internal state coherence \cite{daley_single-atom_2004,griessner_dark-state_2006}. 
In the field of condensed matter simulation, in Bose gases strongly coupled Fr\"ohlich-type polarons have been predicted to form \cite{cucchietti_strong-coupling_2006,*kalas_interaction-induced_2006,*tempere_feynman_2009}. Our system paves the way for experimental studies of multi-polaron systems of one to ten impurities not only in the weak and intermediate coupling regime \cite{casteels_many-polaron_2011}, but also for strong coupling \cite{santamore_multi-impurity_2011}. Adding impurities, and hence polarons, one-by-one would allow to experimentally track the transition even to the many-body regime, and moreover yield information about spatial cluster formation \cite{Klein_Clustering_2007}.  Applied to fermi systems, well localized single spins might allow for realizing model systems to study Kondo physics \cite{Gorshkov_Magnetism_2010}. 
Finally, fundamental questions of quantum physics can be addressed with unprecedented precision, as single impurities can act as local and non-destructive probes of strongly correlated  quantum many-body states \cite{ng_single-atom-aided_2008}. Single atom sensitivity allows to observe single events of molecule formation yielding valuable insight into interaction properties, as we show in this paper, similar to work using single trapped ions \cite{Ratschbacher_ControllingReactions_2012, *Harter_ReactionCenter_2012}. Further, adding single impurities one-by-one to an inititally integrable system, such as a quasi 1D Bose gas \cite{Kinoshita_Cradle_2005}, allows to controllably induce thermalization of a non-equilibrium quantum state.

In addition to tight control over individual impurities, it is a well controlled interaction at ultracold temperatures between quantum gas and impurity which lies at the heart of all these applications in order to preserve the coherence of the sub-systems. In recent realizations of single trapped ions immersed in a BEC \cite{zipkes_trapped_2010,*schmid_dynamics_2010}, mK ion temperatures currently obstruct a coherent time evolution of the hybrid system.

Here, we present the deterministic immersion of single or few neutral Cs atoms in an ultracold Rb gas. We explicitly demonstrate the time-resolved sympathetic cooling of a well-defined number of impurity atoms down to single impurities to the temperature of the many-body system, governed by elastic $s$-wave collisions as required in all scenarios above. As a first application exploiting the properties our hybrid system, we investigate three-body collisions event-by-event with single atom resolution, yielding a precise value of the three-body decay coefficient in Rb-Rb-Cs molecule formation. 

\begin{figure}
    \resizebox{0.4\textwidth}{!}{%
    \includegraphics{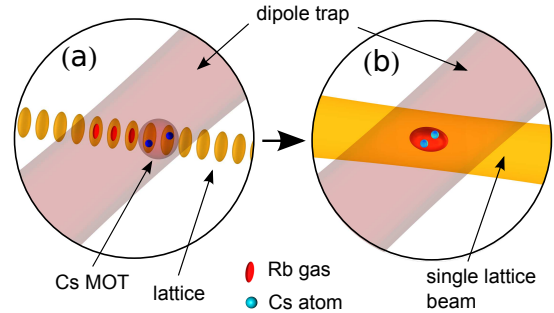}
    }  
  \caption{Schematic setup of the experiment. \textbf{(a)} Rb atoms are precooled in a magnetic trap and then transferred into a crossed dipole trap to a position in close vicinity of a single-atom high gradient magneto-optical trap for Cs. Subsequently, both species are loaded into separated sites of a 1Dl lattice. \textbf{(b)} By adiabatically ramping down the lattice, single Cs atoms are immersed in the ultracold Rb gas stored in the crossed dipole trap.}
\label{setup}
\end{figure}

We use an ultracold gas above the condensation threshold at a high phase-space density of about 0.2. This supports a clear interpretation of the observed interaction effects in the absence of quantum effects. However, the quantum regime will be at the focus of future experiments, and it can be easily accessed by cooling the Rb gas a bit further below the critical temperature.

The insertion of single $^{133}$Cs atoms relies on methods described in Ref.~\cite{spethmann_inserting_2012}. An ultracold $^{87}$Rb gas is produced and stored in a magnetic-field insensitive state ($F = 1$, $m_\mathrm{F} = 0$) in a crossed dipole trap, where $F$ and $m_\mathrm{F}$ are the total angular momentum and its projection onto the quantization axis, respectively. Single Cs atoms are captured in close vicinity in a high gradient magneto-optical trap (MOT). Subsequently, both species are loaded into a one-dimensional (1D) optical lattice at separated lattice sites (Fig. \ref{setup}), and the Cs-MOT is switched off. Here, the Cs atom has a temperature of $\approx$ 30 \textmu K. Finally, the lattice is adiabatically removed while keeping an overall trapping potential by the crossed dipole trap, forcing both species to interact in this common potential. Adiabatic expansion upon removing the lattice cools the Cs atom to a temperature of $\approx$ 5 \textmu K in the 
crossed trap, and the temperature of the 
Rb gas is adjusted to either 250 nK or 700 nK. The ultracold gas is detected by absorption imaging is employed; impurity atoms are counted by monitoring their fluorescence after recapture in the MOT using a single-photon counter. The recapture probability for an atom present in the optical trap is close to unity, so that we assume that the fraction of atoms recaptured is identical to the survival probability.

\begin{figure}
    \resizebox{.5\textwidth}{!}{%
    \includegraphics{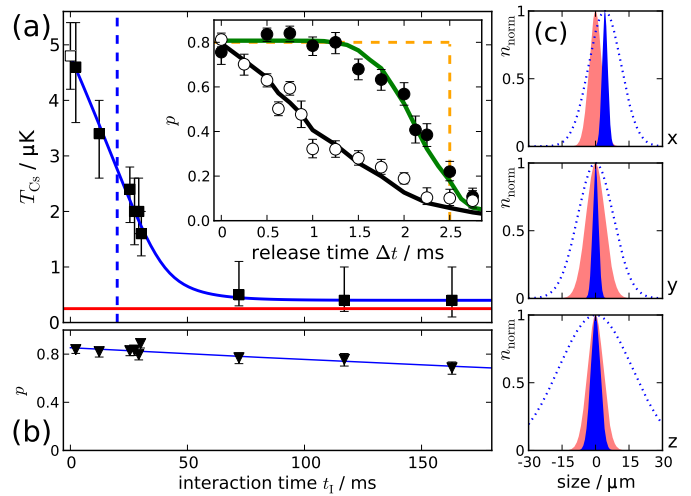}
    }  
  \caption{\textbf{(a)} Cooling of single Cs atoms interacting with an ultracold Rb gas. Within about $50$ ms, Cs is cooled to the temperature of the Rb gas, the solid line shows the cooling according to our model. The horizontal line indicates the Rb temperature $T_\mathrm{Rb} \approx 250$ nK, vertical dashed line indicates the time when the optical lattice is fully extinguished. Inset: Release-recapture measurement for Cs without (open circles) and with 117 ms interaction time with Rb (closed circles). Solid lines indicate the corresponding simulation. The dashed line shows the simplified T=0 recapture curve, see text. \textbf{(b)} Lifetime of Cs during cooling. \textbf{(c)} Normalized density distribution of Rb (light shaded) and Cs before sympathetic cooling (dashed) and after sympathetic cooling (shaded).}
\label{thermalisieren}
\end{figure}

In balanced Rb-Cs mixtures, a large repulsive interspecies interaction was observed, which together with measured Feshbach resonances \cite{pilch_observation_2009} led to a precise knowledge of the two-body molecular interaction potential \cite{anderlini_sympathetic_2005, lercher_production_2011, takekoshi_towards_2012}. A rapid thermalization of the immersed impurity is therefore expected. We study the thermalization of on average two ``hot`` Cs impurity atoms in $F = 3$, interacting via $s$-wave collisions with a large ``cold`` Rb cloud with $N_\mathrm{Rb} \approx$ 12000 in $F = 1$, $m_\mathrm{F} = 0$. As the lattice is removed, the impurity gets in contact with the Rb gas, and the sample is stored in the crossed dipole trap with a potential depth of about 30 \textmu K (70 \textmu K) for Rb (Cs). 

After a variable interaction time $t_\mathrm{I}$, Rb is pushed out of the trap with a resonant light pulse. We have verified that this light pulse does not influence Cs. For each interspecies interaction time $t_\mathrm{I}$, the temperature of Cs is measured with the release-recapture method \cite{mudrich_sympathetic_2002,tuchendler_energy_2008}. 

For this, the trap is switched off diabatically for a release time $\Delta t$, in which Cs is allowed to expand freely. The survival probability after $\Delta t$ depends on the kinetic energy and thus on the temperature: Hot atoms leave the recapture volume faster than cold atoms. The limit of survival is given by gravity which leads to a loss of atoms even at $T=0$  in a time of $\Delta t_\mathrm{l} \approx 2.5$ ms for our trap parameters, see inset of Fig.~\ref{thermalisieren}(a). For atoms at finite temperatures exhibiting an energy distribution, this decay is smeared out towards smaller survival probabilities at shorter $\Delta$t. 

The inset of Fig.~\ref{thermalisieren}(a) shows two examples of release-recapture measurements for two different interspecies interaction times $t_\mathrm{I}$. The survival probability starts at about 80 \%, limited by losses during the loading procedure \cite{spethmann_inserting_2012}. With increasing $\Delta t$ the survival probability decays in a characteristic way. Cs atoms not interacting with Rb show a release half-life time of $\approx 1$\, ms, whereas Cs atoms interacting for 117 ms with Rb feature an almost doubled release half-life time of $\approx 2$\, ms, which indicates cooling.

To quantitatively estimate the temperature of the impurity atoms, the release-recapture experiment is modeled by a numerical simulation. The inset of Fig. \ref{thermalisieren}(a) shows the corresponding simulation that best fits the experimental data. For every interaction time the main graph of Fig. \ref{thermalisieren}(a) shows the extracted temperature. From the initial temperature of $T \approx$ 4.8 \textmu K without interaction, Cs is cooled within about $50$ ms to the temperature of the Rb gas of $T \approx 250$ nK, as determined independently by 
time-of-flight velocimetry. For even longer interaction times, the temperatures of Cs and Rb agree within the uncertainty, which we interpret as thermal equilibrium between the two sub-systems, forming an ideal starting point for the realization of the scenarios listed above. Our system moreover also yields insight  into the non-equilibrium cooling dynamics, where in principle we have access to the full energy distribution at all times.
 
In order to model the sympathetic cooling over the full range, the interspecies density overlap $\bar{n} = \frac{N_\mathrm{Rb} + N_\mathrm{Cs}}{N_\mathrm{Rb}N_\mathrm{Cs}} \int n_\mathrm{Rb}n_\mathrm{Cs}d^3 r $ and the mean relative velocity $\bar{v} = \sqrt{\frac{8k_\mathrm{B}}{\pi}\left(\frac{T_\mathrm{Rb}}{m_\mathrm{Rb}} + \frac{T_\mathrm{Cs}}{m_\mathrm{Cs}}\right)} $ are assumed to be time-dependent. The interspecies scattering cross section $\sigma_\mathrm{RbCs}$ can be calculated from the thermalization by 

\begin{equation}
  -\frac{1}{\Delta T}\frac{d}{dt}\Delta T = \sigma_\mathrm{RbCs}\bar{n}\bar{v}\frac{\xi}{3}
  \label{T_model}
\end{equation}

where $\xi = 0.96$ is the reduction factor due to the mass difference \cite{mudrich_sympathetic_2002}. Fig. \ref{thermalisieren}(a) shows the result of solving Eq. \ref{T_model} numerically for an energy independent scattering cross section $\sigma_\mathrm{RbCs}$, that best fits the experiment. The shape clearly deviates from a purely exponential cooling as usually observed in the thermalization of bulk gases \cite{mudrich_sympathetic_2002,ivanov_sympathetic_2011}, which is a consequence of the negligible perturbation of the Rb gas by the impurity. The corresponding effective $s$-wave scattering length is $|a_\mathrm{RbCs}| = \sqrt{\sigma_\mathrm{RbCs}/4\pi} \approx 450\,a_\mathrm{0}$. 

The $s$-wave scattering lengths for Rb-Cs collisions are known to be $a_\mathrm{S} = \left( 997 \pm 11\right) a_\mathrm{0}$ and $a_\mathrm{T} = \left( 513.3 \pm 2.2\right) a_\mathrm{0}$ for the singlet and triplet potentials, respectively \cite{takekoshi_towards_2012}. Since we only control the hyperfine state of Cs and Rb is in the $F =1$,   $m_\mathrm{F} = 0$-state, $|a_\mathrm{RbCs}|$ is comprised by contributions of scattering channels of the singlet and triplet potentials, so that one would expect $a_\mathrm{T} < |a_\mathrm{RbCs}| < a_\mathrm{S}$.  The discrepancy of our result can be explained by two issues particular to our experiment. Firstly, we load Cs off-center into the trap so that initially the impurity atom could follow a trajectory which has a reduced overlap with the Rb gas. Secondly, the adiabatic lowering of the lattice causes Cs atoms to leave their initial site at non-zero lattice depth, similar to standard evaporation. We indeed observe cooling of Cs already a few ms before 
completely extinguishing the lattice, indicated by the vertical dashed line in Fig. \ref{thermalisieren}(a). These two effects lead to a decreased effective interaction time with respect to $t_\mathrm{I}$. Therefore, the derived value for $|a_\mathrm{RbCs}|$ should be regarded as a lower limit and the sympathetic cooling can be understood in the framework of the well-known two-body collision properties \cite{anderlini_sympathetic_2005, lercher_production_2011, takekoshi_towards_2012}.

The dynamics of our thermalization experiment are remarkable: During sympathetic cooling, the survival probability of impurity atoms decays by a few percent only, as illustrated in Fig. \ref{thermalisieren}(b). Since we precisely know the initial number of impurity atoms, we observe the interaction of individual particles, rather than a statistical ensemble of a bulk gas in imbalanced mixtures \cite{schirotzek_observation_2009,*palzer_quantum_2009,nascimbene_collective_2009,*will_coherent_2011,*trenkwalder_hydrodynamic_2011}. This could be employed, for instance, for the preparation of a mesoscopic number of ultracold or degenerate particles. The heating of the Rb gas, caused by the sympathetic cooling of a single impurity Cs atom, is estimated to be below one nK and can be neglected. Furthermore, the atom number of the Rb gas is approximately constant (discussed below), thus the properties of the Rb gas remain unaffected and the many-body system is not perturbed by the impurity. Cs, in 
contrast, is cooled by more than one order of magnitude in temperature, causing the trapping volume to shrink significantly. In Fig. \ref{thermalisieren}(c), the normalized density distributions of both species before and after thermalization are plotted, illustrating the immersion of the impurity atom into the Rb gas. Thus, our results represent the 
realization of a non-perturbative temperature (and density) probe for a many-body system,  which could be extended to a local and coherent probe. 

While interspecies elastic two-body collisions, that define the properties of our hybrid system, are well understood \cite{anderlini_sympathetic_2005, lercher_production_2011, takekoshi_towards_2012}, interspecies inelastic three-body collisions, limiting the lifetime, are more challenging to handle theoretically and experimentally. The combination of Rb and Cs is particularly involved, as this mixture shows near-resonant scattering properties. Both the Cs-Cs and the Rb-Cs interaction potential feature weakly bound dimer states, which cause a very high three-body recombination rate in both loss channels. Three-body recombination then causes significant heating, leading to non-negligible evaporation loss from the trap (``anti-evaporation`` \cite{weber_three-body_2003}), which in general affects each species to a different level, due to species-dependent trap depths. These effects lead to complicated dynamics, i.e. time-dependent changes in density and temperature of each species, impeding a clear analysis of the three-body processes in (nearly) balanced Rb-Cs mixtures \cite{grimm_private,cho_high_2011,lercher_production_2011}. Additionally, the analysis of losses in bulk gases is further hindered by avalanche effects \cite{schuster_avalanches_2001, *zaccanti_observation_2009}.

Our hybrid system represents a novel approach that is not affected by these effects. We can employ single impurities as individual probes that allow observing inelastic loss mechanisms atom-by-atom and event-by-event. 

\begin{figure}[b!]
    \resizebox{.5\textwidth}{!}{%
    \includegraphics{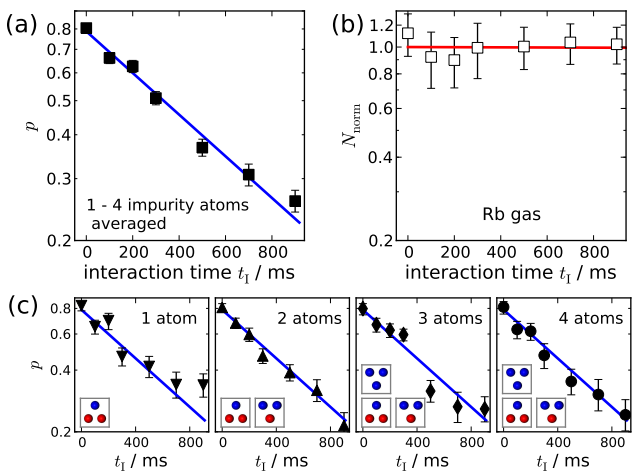}
    }  
  \caption{\textbf{(a)} Decay of on average 1 -- 4 impurity atoms. The solid line is an exponential decay. \textbf{(b)} Rb atom number for the same time interval. The solid line is the theoretical lifetime due to intraspecies three-body decay and collisions with the background gas. \textbf{(c)} Lifetime of a precisely defined number of impurity atoms, post-selected from (a). Solid lines show the same fit as in \textbf{(a)}. 
The insets illustrate the respective open loss channels.}
\label{lifetime}
\end{figure}

Preparing both species in their respective absolute groundstates (Rb: $F=1,m_\mathrm{F}=1$, Cs: $F=3,m_\mathrm{F}=3$) allows for inter- and intraspecies three-body losses, only.  After preparation, both species are stored in a crossed dipole trap at a temperature of $T \approx $ 700 nK, with a potential depth of 54 \textmu K (96 \textmu K) for Rb (Cs). Due to the wavelength of the dipole trap laser, the density distributions are approximately the same for both species, and the differential gravitational sag is negligible for the following experiments. The Rb peak density is 1.2$\times 10^{13}$ cm$^{-3}$ and the three-body loss coefficient is $K_3 = 3 \times 10^{-29}$ cm$^6$s$^{-1}$, so that inelastic intraspecies Rb three-body collisions can be neglected on the relevant time scale, see Fig.~\ref{lifetime}(b). Furthermore, the data shows that the lifetime of the Rb gas is not affected by the impurity atoms, as only few atoms can be lost causing a relative Rb loss of $\approx 10^{-4}$.

For the impurity atoms, Fig. \ref{lifetime}(a) shows the lifetime averaged for 1 -- 4 Cs atoms immersed in the Rb gas, well described by a simple exponential decay with time constant $\tau_\mathrm{t} = (732\pm47)$ ms. The atomic resolution of the impurity atom number allows us to post-select the data for a precisely defined atom number. The corresponding data extracted from the averaged measurement is shown in Fig. \ref{lifetime}(c). For a single impurity atom, the only remaining loss channel is a Rb-Rb-Cs collision. Accordingly, for two impurity atoms also Cs-Cs-Rb collisions are possible, whereas for three and four impurity atoms all loss channels are open. With the same decay as obtained from the average data, the lifetimes of a single, two, three and four atoms are well described, suggesting that Cs-Cs-Rb and Cs-Cs-Cs three-body collisions can be ruled out here. 

\begin{figure}
    \resizebox{.5\textwidth}{!}{%
    \includegraphics{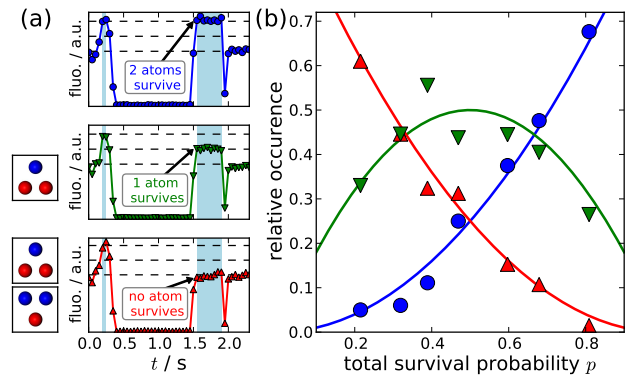}
    }  
  \caption{Three-body losses for exactly two impurity atoms, observed event-by-event. \textbf{(a)} Fluorescence traces showing all three possible loss events and the according allowed loss channels. \textbf{(b)} Corresponding relative occurrences for all possible case: two atom loss (\textcolor{red}{$\blacktriangle$}), one atom loss (\textcolor{darkgreen}{$\blacktriangledown$}) and no atom loss (\textcolor{blue}{\textbullet}). The solid lines show the expectation, for details see text.}
\label{lifetime_stats}
\end{figure}

This can be further verified by analyzing the statistics of the losses, relevant for future studies of two-body interactions. For the case of exactly two impurity atoms immersed in the Rb gas, the respective probabilities $p_\mathrm{i}$, of losing $i = 0,1,2$  Cs atoms are determined for each interaction time $t_\mathrm{I}$, corresponding to a certain total survival probability $p$. Assuming independent impurities and thus the same, independent total survival probability $p$ for each Cs atom, these probabilities are given by $p_\mathrm{0} = p^2$, $p_\mathrm{1} = 2p(1-p)$ and $p_\mathrm{2} = (1-p)(1-p)$. In Fig. \ref{lifetime_stats}, this expectation is plotted together with the corresponding relative occurrences obtained from the experiment, showing good agreement. The impurity atoms are therefore independently interacting with the ultracold Rb gas and hence the observed decay is due to Rb-Rb-Cs three-body recombination. We stress that the ability of studying every single interaction event with single-atom 
resolution is the key for the unambiguous interpretation of inelastic collisions.

Overall, the decay of survival probability of Cs atoms due to three-body collisions with Rb can be described by 

\begin{equation}
 -\frac{1}{N_\mathrm{Cs}}\frac{d}{dt}N_\mathrm{Cs} = L_3 \langle n_\mathrm{Rb}^2  \rangle 
\label{diff_threebody}
\end{equation}

where $ \langle n_\mathrm{Rb}^2 \rangle $ is the mean squared Rb density and $N_\mathrm{Cs}$ the Cs atom number. As described above, the mean squared Rb density $\langle n_\mathrm{Rb}^2 \rangle$ is essentially constant and Eq.~(\ref{diff_threebody}) is simply solved by $N(t) = N_\mathrm{0} \exp \left( -t/\tau_\mathrm{t} \right) = N_\mathrm{0}\exp\left( - L_3 \langle n_\mathrm{Rb}^2 \rangle t\right) $. We obtain the three-body loss coefficient $L_3 = (5\pm2) \times 10 ^{-26} \, \textnormal{cm}^6\textnormal{s}^{-1}$ \cite{Anmerkung1}, in very good agreement with a recent theoretical calculation \cite{wang_private}. Our approach allows therefore a precise determination of $L_3$, so far impossible with balanced mixtures \cite{cho_high_2011,lercher_production_2011}.

In conclusion, our experiment demonstrates the creation of a novel hybrid system, comprised of an atomic impurity interacting with a many-body system via $s$-wave collisions. Recently found Feshbach resonances \cite{pilch_observation_2009} will allow us tuning the interaction of the two subsystems in future experiments, while the use of species selective potentials \cite{leblanc_speciespecific_2007} will improve localization of impurities without perturbing the gas. Together with site-resolved detection beyond the diffraction limit \cite{karski_nearest-neighbor_2009}, this provides a good starting point for the realization of the scenarios mentioned above.

We gratefully acknowledge partial support from the science ministry of North Rhine-Westfalia (NRW MWIFT) through an independent Junior Research Group. N.S. acknowledges support from Studienstiftung des Deutschen Volkes and N.S. and C.W. from the Bonn-Cologne Graduate School of Physics and Astronomy.
 

\begin{thebibliography}{10}%
\makeatletter
\providecommand \@ifxundefined [1]{%
 \ifx #1\undefined \expandafter \@firstoftwo
 \else \expandafter \@secondoftwo
\fi
}%
\providecommand \@ifnum [1]{%
 \ifnum #1\expandafter \@firstoftwo
 \else \expandafter \@secondoftwo
\fi
}%
\providecommand \enquote [1]{``#1''}%
\providecommand \bibnamefont  [1]{#1}%
\providecommand \bibfnamefont [1]{#1}%
\providecommand \citenamefont [1]{#1}%
\providecommand\href[0]{\@sanitize\@href}%
\providecommand\@href[1]{\endgroup\@@startlink{#1}\endgroup\@@href}%
\providecommand\@@href[1]{#1\@@endlink}%
\providecommand \@sanitize [0]{\begingroup\catcode`\&12\catcode`\#12\relax}%
\@ifxundefined \pdfoutput {\@firstoftwo}{%
 \@ifnum{\z@=\pdfoutput}{\@firstoftwo}{\@secondoftwo}%
}{%
 \providecommand\@@startlink[1]{\leavevmode}%
 \providecommand\@@endlink[0]{}%
}{%
 \providecommand\@@startlink[1]{%
  \leavevmode
  \pdfstartlink
   attr{/Border[0 0 1 ]/H/I/C[0 1 1]}%
   user{/Subtype/Link/A<</Type/Action/S/URI/URI(#1)>>}%
  \relax
 }%
 \providecommand\@@endlink[0]{\pdfendlink}%
}%
\providecommand \url  [0]{\begingroup\@sanitize \@url }%
\providecommand \@url [1]{\endgroup\@href {#1}{\urlprefix}}%
\providecommand \urlprefix [0]{URL }%
\providecommand \Eprint[0]{\href }%
\@ifxundefined \urlstyle {%
  \providecommand \doi [1]{doi:\discretionary{}{}{}#1}%
}{%
  \providecommand \doi [0]{doi:\discretionary{}{}{}\begingroup
  \urlstyle{rm}\Url }%
}%
\providecommand \doibase [0]{http://dx.doi.org/}%
\providecommand \Doi[1]{\href{\doibase#1}}%
\providecommand \bibAnnote [3]{%
  \BibitemShut{#1}%
  \begin{quotation}\noindent
    \textsc{Key:}\ #2\\\textsc{Annotation:}\ #3%
  \end{quotation}%
}%
\providecommand \bibAnnoteFile [2]{%
  \IfFileExists{#2}{\bibAnnote {#1} {#2} {\input{#2}}}{}%
}%
\providecommand \typeout [0]{\immediate \write \m@ne }%
\providecommand \selectlanguage [0]{\@gobble}%
\providecommand \bibinfo [0]{\@secondoftwo}%
\providecommand \bibfield [0]{\@secondoftwo}%
\providecommand \translation [1]{[#1]}%
\providecommand \BibitemOpen[0]{}%
\providecommand \bibitemStop [0]{}%
\providecommand \bibitemNoStop [0]{.\EOS\space}%
\providecommand \EOS [0]{\spacefactor3000\relax}%
\providecommand \BibitemShut [1]{\csname bibitem#1\endcsname}%
\bibitem{gericke_high-resolution_2008}%
  \BibitemOpen
  \bibfield{author}{%
  \bibinfo {author} {\bibfnamefont{T.}~\bibnamefont{Gericke}}, \bibinfo
  {author} {\bibfnamefont{P.}~\bibnamefont{W\"{u}rtz}}, \bibinfo {author}
  {\bibfnamefont{D.}~\bibnamefont{Reitz}}, \bibinfo {author}
  {\bibfnamefont{T.}~\bibnamefont{Langen}},\ and\ \bibinfo {author}
  {\bibfnamefont{H.}~\bibnamefont{Ott}},\ }%
  \bibfield{journal}{%
  \Doi{10.1038/nphys1102}{\bibinfo {journal} {Nat. Phys.}}\ }%
  \textbf{\bibinfo {volume} {4}},\ \bibinfo {pages} {949} (\bibinfo {year}
  {2008})%
  \bibAnnoteFile{NoStop}{gericke_high-resolution_2008}%
\bibitem{bakr_quantum_2009}%
  \BibitemOpen
  \bibfield{author}{%
  \bibinfo {author} {\bibfnamefont{W.~S.}\ \bibnamefont{Bakr}}, \bibinfo
  {author} {\bibfnamefont{J.~I.}\ \bibnamefont{Gillen}}, \bibinfo {author}
  {\bibfnamefont{A.}~\bibnamefont{Peng}}, \bibinfo {author}
  {\bibfnamefont{S.}~\bibnamefont{F\"{o}lling}},\ and\ \bibinfo {author}
  {\bibfnamefont{M.}~\bibnamefont{Greiner}},\ }%
  \bibfield{journal}{%
  \Doi{10.1038/nature08482}{\bibinfo {journal} {Nature}}\ }%
  \textbf{\bibinfo {volume} {462}},\ \bibinfo {pages} {74} (\bibinfo {year}
  {2009})%
  \bibAnnoteFile{NoStop}{bakr_quantum_2009}%
\bibitem{sherson_single-atom-resolved_2010}%
  \BibitemOpen
  \bibfield{author}{%
  \bibinfo {author} {\bibfnamefont{J.~F.}\ \bibnamefont{Sherson}}, \bibinfo
  {author} {\bibfnamefont{C.}~\bibnamefont{Weitenberg}}, \bibinfo {author}
  {\bibfnamefont{M.}~\bibnamefont{Endres}}, \bibinfo {author}
  {\bibfnamefont{M.}~\bibnamefont{Cheneau}}, \bibinfo {author}
  {\bibfnamefont{I.}~\bibnamefont{Bloch}},\ and\ \bibinfo {author}
  {\bibfnamefont{S.}~\bibnamefont{Kuhr}},\ }%
  \bibfield{journal}{%
  \Doi{10.1038/nature09378}{\bibinfo {journal} {Nature}}\ }%
  \textbf{\bibinfo {volume} {467}},\ \bibinfo {pages} {68} (\bibinfo {year}
  {2010})%
  \bibAnnoteFile{NoStop}{sherson_single-atom-resolved_2010}%
\bibitem{serwane_deterministic_2011}%
  \BibitemOpen
  \bibfield{author}{%
  \bibinfo {author} {\bibfnamefont{F.}~\bibnamefont{Serwane}}, \bibinfo
  {author} {\bibfnamefont{G.}~\bibnamefont{Z\"{u}rn}}, \bibinfo {author}
  {\bibfnamefont{T.}~\bibnamefont{Lompe}}, \bibinfo {author}
  {\bibfnamefont{T.~B.}\ \bibnamefont{Ottenstein}}, \bibinfo {author}
  {\bibfnamefont{A.~N.}\ \bibnamefont{Wenz}},\ and\ \bibinfo {author}
  {\bibfnamefont{S.}~\bibnamefont{Jochim}},\ }%
  \bibfield{journal}{%
  \bibinfo {journal} {Science}\ }%
  \textbf{\bibinfo {volume} {332}},\ \bibinfo {pages} {336 } (\bibinfo {year}
  {2011})%
  \bibAnnoteFile{NoStop}{serwane_deterministic_2011}%
\bibitem{schirotzek_observation_2009}%
  \BibitemOpen
  \bibfield{author}{%
  \bibinfo {author} {\bibfnamefont{A.}~\bibnamefont{Schirotzek}}, \bibinfo
  {author} {\bibfnamefont{C.}~\bibnamefont{Wu}}, \bibinfo {author}
  {\bibfnamefont{A.}~\bibnamefont{Sommer}},\ and\ \bibinfo {author}
  {\bibfnamefont{M.~W.}\ \bibnamefont{Zwierlein}},\ }%
  \bibfield{journal}{%
  \bibinfo {journal} {Phys. Rev. Lett.}\ }%
  \textbf{\bibinfo {volume} {102}},\ \bibinfo {pages} {230402} (\bibinfo {year}
  {2009})%
  \bibAnnoteFile{NoStop}{schirotzek_observation_2009}%
\bibitem{palzer_quantum_2009}%
  \BibitemOpen
  \bibfield{author}{%
  \bibinfo {author} {\bibfnamefont{S.}~\bibnamefont{Palzer}}, \bibinfo {author}
  {\bibfnamefont{C.}~\bibnamefont{Zipkes}}, \bibinfo {author}
  {\bibfnamefont{C.}~\bibnamefont{Sias}},\ and\ \bibinfo {author}
  {\bibfnamefont{M.}~\bibnamefont{K\"{o}hl}},\ }%
  \bibfield{journal}{%
  \Doi{10.1103/PhysRevLett.103.150601}{\bibinfo {journal} {Phys. Rev. Lett.}}\
  }%
  \textbf{\bibinfo {volume} {103}},\ \bibinfo {pages} {150601} (\bibinfo {year}
  {2009})%
  \bibAnnoteFile{NoStop}{palzer_quantum_2009}%
\bibitem{micheli_single_2004}%
  \BibitemOpen
  \bibfield{author}{%
  \bibinfo {author} {\bibfnamefont{A.}~\bibnamefont{Micheli}}, \bibinfo
  {author} {\bibfnamefont{A.~J.}\ \bibnamefont{Daley}}, \bibinfo {author}
  {\bibfnamefont{D.}~\bibnamefont{Jaksch}},\ and\ \bibinfo {author}
  {\bibfnamefont{P.}~\bibnamefont{Zoller}},\ }%
  \bibfield{journal}{%
  \bibinfo {journal} {Phys. Rev. Lett.}\ }%
  \textbf{\bibinfo {volume} {93}},\ \bibinfo {pages} {140408} (\bibinfo {year}
  {2004})%
  \bibAnnoteFile{NoStop}{micheli_single_2004}%
\bibitem{klein_interaction_2005}%
  \BibitemOpen
  \bibfield{author}{%
  \bibinfo {author} {\bibfnamefont{A.}~\bibnamefont{Klein}}\ and\ \bibinfo
  {author} {\bibfnamefont{M.}~\bibnamefont{Fleischhauer}},\ }%
  \bibfield{journal}{%
  \bibinfo {journal} {Phys. Rev. A}\ }%
  \textbf{\bibinfo {volume} {71}},\ \bibinfo {pages} {033605} (\bibinfo {year}
  {2005})%
  \bibAnnoteFile{NoStop}{klein_interaction_2005}%
\bibitem{daley_single-atom_2004}%
  \BibitemOpen
  \bibfield{author}{%
  \bibinfo {author} {\bibfnamefont{A.~J.}\ \bibnamefont{Daley}}, \bibinfo
  {author} {\bibfnamefont{P.~O.}\ \bibnamefont{Fedichev}},\ and\ \bibinfo
  {author} {\bibfnamefont{P.}~\bibnamefont{Zoller}},\ }%
  \bibfield{journal}{%
  \Doi{10.1103/PhysRevA.69.022306}{\bibinfo {journal} {Phys. Rev. A}}\ }%
  \textbf{\bibinfo {volume} {69}},\ \bibinfo {pages} {022306} (\bibinfo {year}
  {2004})%
  \bibAnnoteFile{NoStop}{daley_single-atom_2004}%
\bibitem{griessner_dark-state_2006}%
  \BibitemOpen
  \bibfield{author}{%
  \bibinfo {author} {\bibfnamefont{A.}~\bibnamefont{Griessner}}, \bibinfo
  {author} {\bibfnamefont{A.~J.}\ \bibnamefont{Daley}}, \bibinfo {author}
  {\bibfnamefont{S.~R.}\ \bibnamefont{Clark}}, \bibinfo {author}
  {\bibfnamefont{D.}~\bibnamefont{Jaksch}},\ and\ \bibinfo {author}
  {\bibfnamefont{P.}~\bibnamefont{Zoller}},\ }%
  \bibfield{journal}{%
  \Doi{10.1103/PhysRevLett.97.220403}{\bibinfo {journal} {Phys. Rev. Lett.}}\
  }%
  \textbf{\bibinfo {volume} {97}},\ \bibinfo {pages} {220403} (\bibinfo {year}
  {2006})%
  \bibAnnoteFile{NoStop}{griessner_dark-state_2006}%
\bibitem{cucchietti_strong-coupling_2006}%
  \BibitemOpen
  \bibfield{author}{%
  \bibinfo {author} {\bibfnamefont{F.~M.}\ \bibnamefont{Cucchietti}}\ and\
  \bibinfo {author} {\bibfnamefont{E.}~\bibnamefont{Timmermans}},\ }%
  \bibfield{journal}{%
  \bibinfo {journal} {Phys. Rev. Lett.}\ }%
  \textbf{\bibinfo {volume} {96}},\ \bibinfo {pages} {210401} (\bibinfo {year}
  {2006})%
  \bibAnnoteFile{NoStop}{cucchietti_strong-coupling_2006}%
\bibitem{kalas_interaction-induced_2006}%
  \BibitemOpen
  \bibfield{author}{%
  \bibinfo {author} {\bibfnamefont{R.~M.}\ \bibnamefont{Kalas}}\ and\ \bibinfo
  {author} {\bibfnamefont{D.}~\bibnamefont{Blume}},\ }%
  \bibfield{journal}{%
  \bibinfo {journal} {Phys. Rev. A}\ }%
  \textbf{\bibinfo {volume} {73}},\ \bibinfo {pages} {043608} (\bibinfo {year}
  {2006})%
  \bibAnnoteFile{NoStop}{kalas_interaction-induced_2006}%
\bibitem{tempere_feynman_2009}%
  \BibitemOpen
  \bibfield{author}{%
  \bibinfo {author} {\bibfnamefont{J.}~\bibnamefont{Tempere}}, \bibinfo
  {author} {\bibfnamefont{W.}~\bibnamefont{Casteels}}, \bibinfo {author}
  {\bibfnamefont{M.~K.}\ \bibnamefont{Oberthaler}}, \bibinfo {author}
  {\bibfnamefont{S.}~\bibnamefont{Knoop}}, \bibinfo {author}
  {\bibfnamefont{E.}~\bibnamefont{Timmermans}},\ and\ \bibinfo {author}
  {\bibfnamefont{J.~T.}\ \bibnamefont{Devreese}},\ }%
  \bibfield{journal}{%
  \Doi{10.1103/PhysRevB.80.184504}{\bibinfo {journal} {Phys. Rev. B}}\ }%
  \textbf{\bibinfo {volume} {80}},\ \bibinfo {pages} {184504} (\bibinfo {year}
  {2009})%
  \bibAnnoteFile{NoStop}{tempere_feynman_2009}%
\bibitem{casteels_many-polaron_2011}%
  \BibitemOpen
  \bibfield{author}{%
  \bibinfo {author} {\bibfnamefont{W.}~\bibnamefont{Casteels}}, \bibinfo
  {author} {\bibfnamefont{J.}~\bibnamefont{Tempere}},\ and\ \bibinfo {author}
  {\bibfnamefont{J.~T.}\ \bibnamefont{Devreese}},\ }%
  \bibfield{journal}{%
  \bibinfo {journal} {Phys. Rev. A}\ }%
  \textbf{\bibinfo {volume} {84}},\ \bibinfo {pages} {063612} (\bibinfo {year}
  {2011})%
  \bibAnnoteFile{NoStop}{casteels_many-polaron_2011}%
\bibitem{santamore_multi-impurity_2011}%
  \BibitemOpen
  \bibfield{author}{%
  \bibinfo {author} {\bibfnamefont{D.~H.}\ \bibnamefont{Santamore}}\ and\
  \bibinfo {author} {\bibfnamefont{E.}~\bibnamefont{Timmermans}},\ }%
  \bibfield{journal}{%
  \bibinfo {journal} {New J. Phys.}\ }%
  \textbf{\bibinfo {volume} {13}},\ \bibinfo {pages} {103029} (\bibinfo {year}
  {2011})%
  \bibAnnoteFile{NoStop}{santamore_multi-impurity_2011}%
\bibitem{Klein_Clustering_2007}%
  \BibitemOpen
  \bibfield{author}{%
  \bibinfo {author} {\bibfnamefont{A.}~\bibnamefont{Klein}}, \bibinfo {author}
  {\bibfnamefont{M.}~\bibnamefont{Bruderer}}, \bibinfo {author}
  {\bibfnamefont{C.~S.}\ \bibnamefont{R.}},\ and\ \bibinfo {author}
  {\bibfnamefont{D.}~\bibnamefont{Jaksch}},\ }%
  \bibfield{journal}{%
  \bibinfo {journal} {New J. Phys.}\ }%
  \textbf{\bibinfo {volume} {9}} (\bibinfo {year} {2007})%
  \bibAnnoteFile{NoStop}{Klein_Clustering_2007}%
\bibitem{Gorshkov_Magnetism_2010}%
  \BibitemOpen
  \bibfield{author}{%
  \bibinfo {author} {\bibfnamefont{A.~V.}\ \bibnamefont{Gorshkov}}, \bibinfo
  {author} {\bibfnamefont{M.}~\bibnamefont{Hermele}}, \bibinfo {author}
  {\bibfnamefont{V.}~\bibnamefont{Gurarie}}, \bibinfo {author}
  {\bibfnamefont{C.}~\bibnamefont{Xu}}, \bibinfo {author}
  {\bibfnamefont{P.}~\bibnamefont{Julienne}}, \bibinfo {author}
  {\bibfnamefont{J.}~\bibnamefont{Ye}}, \bibinfo {author}
  {\bibfnamefont{P.}~\bibnamefont{Zoller}}, \bibinfo {author}
  {\bibfnamefont{E.}~\bibnamefont{Demler}}, \bibinfo {author}
  {\bibfnamefont{M.~D.}\ \bibnamefont{Lukin}},\ and\ \bibinfo {author}
  {\bibfnamefont{A.}~\bibnamefont{Rey}},\ }%
  \bibfield{journal}{%
  \bibinfo {journal} {Nature Phys.}\ }%
  \textbf{\bibinfo {volume} {6}} (\bibinfo {year} {2010})%
  \bibAnnoteFile{NoStop}{Gorshkov_Magnetism_2010}%
\bibitem{ng_single-atom-aided_2008}%
  \BibitemOpen
  \bibfield{author}{%
  \bibinfo {author} {\bibfnamefont{H.~T.}\ \bibnamefont{Ng}}\ and\ \bibinfo
  {author} {\bibfnamefont{S.}~\bibnamefont{Bose}},\ }%
  \bibfield{journal}{%
  \Doi{10.1103/PhysRevA.78.023610}{\bibinfo {journal} {Phys. Rev. A}}\ }%
  \textbf{\bibinfo {volume} {78}},\ \bibinfo {pages} {023610} (\bibinfo {year}
  {2008})%
  \bibAnnoteFile{NoStop}{ng_single-atom-aided_2008}%
\bibitem{Ratschbacher_ControllingReactions_2012}%
  \BibitemOpen
  \bibfield{author}{%
  \bibinfo {author} {\bibfnamefont{L.}~\bibnamefont{Ratschbacher}}, \bibinfo
  {author} {\bibfnamefont{C.}~\bibnamefont{Zipkes}}, \bibinfo {author}
  {\bibfnamefont{C.}~\bibnamefont{Sias}},\ and\ \bibinfo {author}
  {\bibfnamefont{M.}~\bibnamefont{K{\"o}hl}},\ }%
  \bibfield{journal}{%
  \bibinfo {journal} {Nature Phys.}\ }%
  \textbf{\bibinfo {volume} {8}} (\bibinfo {year} {2012})%
  \bibAnnoteFile{NoStop}{Ratschbacher_ControllingReactions_2012}%
\bibitem{Harter_ReactionCenter_2012}%
  \BibitemOpen
  \bibfield{author}{%
  \bibinfo {author} {\bibfnamefont{A.}~\bibnamefont{H\"arter}}, \bibinfo
  {author} {\bibfnamefont{A.}~\bibnamefont{Kr\"ukow}}, \bibinfo {author}
  {\bibfnamefont{A.}~\bibnamefont{Brunner}}, \bibinfo {author}
  {\bibfnamefont{W.}~\bibnamefont{Schnitzler}}, \bibinfo {author}
  {\bibfnamefont{S.}~\bibnamefont{Schmid}},\ and\ \bibinfo {author}
  {\bibfnamefont{J.}~\bibnamefont{Hecker~Denschlag}},\ }%
  \bibfield{journal}{%
  \bibinfo {journal} {arXiv:1206.0242}}%
   (\bibinfo {year} {2012})%
  \bibAnnoteFile{NoStop}{Harter_ReactionCenter_2012}%
\bibitem{Kinoshita_Cradle_2005}%
  \BibitemOpen
  \bibfield{author}{%
  \bibinfo {author} {\bibfnamefont{T.}~\bibnamefont{Kinoshita}}, \bibinfo
  {author} {\bibfnamefont{W.}~\bibnamefont{Trevor}},\ and\ \bibinfo {author}
  {\bibfnamefont{D.~S.}\ \bibnamefont{Weiss}},\ }%
  \bibfield{journal}{%
  \bibinfo {journal} {Nature}\ }%
  \textbf{\bibinfo {volume} {440}} (\bibinfo {year} {2005})%
  \bibAnnoteFile{NoStop}{Kinoshita_Cradle_2005}%
\bibitem{zipkes_trapped_2010}%
  \BibitemOpen
  \bibfield{author}{%
  \bibinfo {author} {\bibfnamefont{C.}~\bibnamefont{Zipkes}}, \bibinfo {author}
  {\bibfnamefont{S.}~\bibnamefont{Palzer}}, \bibinfo {author}
  {\bibfnamefont{C.}~\bibnamefont{Sias}},\ and\ \bibinfo {author}
  {\bibfnamefont{M.}~\bibnamefont{K\"{o}hl}},\ }%
  \bibfield{journal}{%
  \Doi{10.1038/nature08865}{\bibinfo {journal} {Nature}}\ }%
  \textbf{\bibinfo {volume} {464}},\ \bibinfo {pages} {388} (\bibinfo {year}
  {2010})%
  \bibAnnoteFile{NoStop}{zipkes_trapped_2010}%
\bibitem{schmid_dynamics_2010}%
  \BibitemOpen
  \bibfield{author}{%
  \bibinfo {author} {\bibfnamefont{S.}~\bibnamefont{Schmid}}, \bibinfo {author}
  {\bibfnamefont{A.}~\bibnamefont{H\"{a}rter}},\ and\ \bibinfo {author}
  {\bibfnamefont{J.}~\bibnamefont{Hecker~Denschlag}},\ }%
  \bibfield{journal}{%
  \Doi{10.1103/PhysRevLett.105.133202}{\bibinfo {journal} {Phys. Rev. Lett.}}\
  }%
  \textbf{\bibinfo {volume} {105}},\ \bibinfo {pages} {133202} (\bibinfo {year}
  {2010})%
  \bibAnnoteFile{NoStop}{schmid_dynamics_2010}%
\bibitem{spethmann_inserting_2012}%
  \BibitemOpen
  \bibfield{author}{%
  \bibinfo {author} {\bibfnamefont{N.}~\bibnamefont{Spethmann}}, \bibinfo
  {author} {\bibfnamefont{F.}~\bibnamefont{Kindermann}}, \bibinfo {author}
  {\bibfnamefont{S.}~\bibnamefont{John}}, \bibinfo {author}
  {\bibfnamefont{C.}~\bibnamefont{Weber}}, \bibinfo {author}
  {\bibfnamefont{D.}~\bibnamefont{Meschede}},\ and\ \bibinfo {author}
  {\bibfnamefont{A.}~\bibnamefont{Widera}},\ }%
  \bibfield{journal}{%
  \bibinfo {journal} {Appl. Phys. B}\ }%
  \textbf{\bibinfo {volume} {106}},\ \bibinfo {pages} {513} (\bibinfo {year}
  {2012})%
  \bibAnnoteFile{NoStop}{spethmann_inserting_2012}%
\bibitem{pilch_observation_2009}%
  \BibitemOpen
  \bibfield{author}{%
  \bibinfo {author} {\bibfnamefont{K.}~\bibnamefont{Pilch}}, \bibinfo {author}
  {\bibfnamefont{A.~D.}\ \bibnamefont{Lange}}, \bibinfo {author}
  {\bibfnamefont{A.}~\bibnamefont{Prantner}}, \bibinfo {author}
  {\bibfnamefont{G.}~\bibnamefont{Kerner}}, \bibinfo {author}
  {\bibfnamefont{F.}~\bibnamefont{Ferlaino}}, \bibinfo {author}
  {\bibfnamefont{H.}~\bibnamefont{N\"{a}gerl}},\ and\ \bibinfo {author}
  {\bibfnamefont{R.}~\bibnamefont{Grimm}},\ }%
  \bibfield{journal}{%
  \bibinfo {journal} {Phys. Rev. A}\ }%
  \textbf{\bibinfo {volume} {79}},\ \bibinfo {pages} {042718} (\bibinfo {year}
  {2009})%
  \bibAnnoteFile{NoStop}{pilch_observation_2009}%
\bibitem{anderlini_sympathetic_2005}%
  \BibitemOpen
  \bibfield{author}{%
  \bibinfo {author} {\bibfnamefont{M.}~\bibnamefont{Anderlini}}, \bibinfo
  {author} {\bibfnamefont{E.}~\bibnamefont{Courtade}}, \bibinfo {author}
  {\bibfnamefont{M.}~\bibnamefont{Cristiani}}, \bibinfo {author}
  {\bibfnamefont{D.}~\bibnamefont{Cossart}}, \bibinfo {author}
  {\bibfnamefont{D.}~\bibnamefont{Ciampini}}, \bibinfo {author}
  {\bibfnamefont{C.}~\bibnamefont{Sias}}, \bibinfo {author}
  {\bibfnamefont{O.}~\bibnamefont{Morsch}},\ and\ \bibinfo {author}
  {\bibfnamefont{E.}~\bibnamefont{Arimondo}},\ }%
  \bibfield{journal}{%
  \Doi{10.1103/PhysRevA.71.061401}{\bibinfo {journal} {Phys. Rev. A}}\ }%
  \textbf{\bibinfo {volume} {71}},\ \bibinfo {pages} {061401} (\bibinfo {year}
  {2005})%
  \bibAnnoteFile{NoStop}{anderlini_sympathetic_2005}%
\bibitem{lercher_production_2011}%
  \BibitemOpen
  \bibfield{author}{%
  \bibinfo {author} {\bibfnamefont{A.~D.}\ \bibnamefont{Lercher}}, \bibinfo
  {author} {\bibfnamefont{T.}~\bibnamefont{Takekoshi}}, \bibinfo {author}
  {\bibfnamefont{M.}~\bibnamefont{Debatin}}, \bibinfo {author}
  {\bibfnamefont{B.}~\bibnamefont{Schuster}}, \bibinfo {author}
  {\bibfnamefont{R.}~\bibnamefont{Rameshan}}, \bibinfo {author}
  {\bibfnamefont{F.}~\bibnamefont{Ferlaino}}, \bibinfo {author}
  {\bibfnamefont{R.}~\bibnamefont{Grimm}},\ and\ \bibinfo {author}
  {\bibfnamefont{H.}~\bibnamefont{N\"{a}gerl}},\ }%
  \bibfield{journal}{%
  \bibinfo {journal} {Euro. Phys. Jour. D}\ }%
  \textbf{\bibinfo {volume} {65}},\ \bibinfo {pages} {3} (\bibinfo {year}
  {2011})%
  \bibAnnoteFile{NoStop}{lercher_production_2011}%
\bibitem{takekoshi_towards_2012}%
  \BibitemOpen
  \bibfield{author}{%
  \bibinfo {author} {\bibfnamefont{T.}~\bibnamefont{Takekoshi}}, \bibinfo
  {author} {\bibfnamefont{M.}~\bibnamefont{Debatin}}, \bibinfo {author}
  {\bibfnamefont{R.}~\bibnamefont{Rameshan}}, \bibinfo {author}
  {\bibfnamefont{F.}~\bibnamefont{Ferlaino}}, \bibinfo {author}
  {\bibfnamefont{R.}~\bibnamefont{Grimm}}, \bibinfo {author}
  {\bibfnamefont{H.}~\bibnamefont{N\"{a}gerl}}, \bibinfo {author}
  {\bibfnamefont{C.~R.}\ \bibnamefont{Le~Sueur}}, \bibinfo {author}
  {\bibfnamefont{J.~M.}\ \bibnamefont{Hutson}}, \bibinfo {author}
  {\bibfnamefont{P.~S.}\ \bibnamefont{Julienne}}, \bibinfo {author}
  {\bibfnamefont{S.}~\bibnamefont{Kotochigova}},\ and\ \bibinfo {author}
  {\bibfnamefont{E.}~\bibnamefont{Tiemann}},\ }%
  \bibfield{journal}{%
  \bibinfo {journal} {Phys. Rev. A}\ }%
  \textbf{\bibinfo {volume} {85}},\ \bibinfo {pages} {032506} (\bibinfo {year}
  {2012})%
  \bibAnnoteFile{NoStop}{takekoshi_towards_2012}%
\bibitem{mudrich_sympathetic_2002}%
  \BibitemOpen
  \bibfield{author}{%
  \bibinfo {author} {\bibfnamefont{M.}~\bibnamefont{Mudrich}}, \bibinfo
  {author} {\bibfnamefont{S.}~\bibnamefont{Kraft}}, \bibinfo {author}
  {\bibfnamefont{K.}~\bibnamefont{Singer}}, \bibinfo {author}
  {\bibfnamefont{R.}~\bibnamefont{Grimm}}, \bibinfo {author}
  {\bibfnamefont{A.}~\bibnamefont{Mosk}},\ and\ \bibinfo {author}
  {\bibfnamefont{M.}~\bibnamefont{Weidem\"{u}ller}},\ }%
  \bibfield{journal}{%
  \Doi{10.1103/PhysRevLett.88.253001}{\bibinfo {journal} {Phys. Rev. Lett.}}\
  }%
  \textbf{\bibinfo {volume} {88}},\ \bibinfo {pages} {253001} (\bibinfo {year}
  {2002})%
  \bibAnnoteFile{NoStop}{mudrich_sympathetic_2002}%
\bibitem{tuchendler_energy_2008}%
  \BibitemOpen
  \bibfield{author}{%
  \bibinfo {author} {\bibfnamefont{C.}~\bibnamefont{Tuchendler}}, \bibinfo
  {author} {\bibfnamefont{A.~M.}\ \bibnamefont{Lance}}, \bibinfo {author}
  {\bibfnamefont{A.}~\bibnamefont{Browaeys}}, \bibinfo {author}
  {\bibfnamefont{Y.~R.~P.}\ \bibnamefont{Sortais}},\ and\ \bibinfo {author}
  {\bibfnamefont{P.}~\bibnamefont{Grangier}},\ }%
  \bibfield{journal}{%
  \Doi{10.1103/PhysRevA.78.033425}{\bibinfo {journal} {Phys. Rev. A}}\ }%
  \textbf{\bibinfo {volume} {78}},\ \bibinfo {pages} {033425} (\bibinfo {year}
  {2008})%
  \bibAnnoteFile{NoStop}{tuchendler_energy_2008}%
\bibitem{ivanov_sympathetic_2011}%
  \BibitemOpen
  \bibfield{author}{%
  \bibinfo {author} {\bibfnamefont{V.~V.}\ \bibnamefont{Ivanov}}, \bibinfo
  {author} {\bibfnamefont{A.}~\bibnamefont{Khramov}}, \bibinfo {author}
  {\bibfnamefont{A.~H.}\ \bibnamefont{Hansen}}, \bibinfo {author}
  {\bibfnamefont{W.~H.}\ \bibnamefont{Dowd}}, \bibinfo {author}
  {\bibfnamefont{F.}~\bibnamefont{M\"{u}nchow}}, \bibinfo {author}
  {\bibfnamefont{A.~O.}\ \bibnamefont{Jamison}},\ and\ \bibinfo {author}
  {\bibfnamefont{S.}~\bibnamefont{Gupta}},\ }%
  \bibfield{journal}{%
  \bibinfo {journal} {Phys. Rev. Lett.}\ }%
  \textbf{\bibinfo {volume} {106}},\ \bibinfo {pages} {153201} (\bibinfo {year}
  {2011})%
  \bibAnnoteFile{NoStop}{ivanov_sympathetic_2011}%
\bibitem{nascimbene_collective_2009}%
  \BibitemOpen
  \bibfield{author}{%
  \bibinfo {author} {\bibfnamefont{S.}~\bibnamefont{Nascimb\`{e}ne}}, \bibinfo
  {author} {\bibfnamefont{N.}~\bibnamefont{Navon}}, \bibinfo {author}
  {\bibfnamefont{K.~J.}\ \bibnamefont{Jiang}}, \bibinfo {author}
  {\bibfnamefont{L.}~\bibnamefont{Tarruell}}, \bibinfo {author}
  {\bibfnamefont{M.}~\bibnamefont{Teichmann}}, \bibinfo {author}
  {\bibfnamefont{J.}~\bibnamefont{{McKeever}}}, \bibinfo {author}
  {\bibfnamefont{F.}~\bibnamefont{Chevy}},\ and\ \bibinfo {author}
  {\bibfnamefont{C.}~\bibnamefont{Salomon}},\ }%
  \bibfield{journal}{%
  \bibinfo {journal} {Phys. Rev. Lett.}\ }%
  \textbf{\bibinfo {volume} {103}},\ \bibinfo {pages} {170402} (\bibinfo {year}
  {2009})%
  \bibAnnoteFile{NoStop}{nascimbene_collective_2009}%
\bibitem{will_coherent_2011}%
  \BibitemOpen
  \bibfield{author}{%
  \bibinfo {author} {\bibfnamefont{S.}~\bibnamefont{Will}}, \bibinfo {author}
  {\bibfnamefont{T.}~\bibnamefont{Best}}, \bibinfo {author}
  {\bibfnamefont{S.}~\bibnamefont{Braun}}, \bibinfo {author}
  {\bibfnamefont{U.}~\bibnamefont{Schneider}},\ and\ \bibinfo {author}
  {\bibfnamefont{I.}~\bibnamefont{Bloch}},\ }%
  \bibfield{journal}{%
  \bibinfo {journal} {Phys. Rev. Lett.}\ }%
  \textbf{\bibinfo {volume} {106}},\ \bibinfo {pages} {115305} (\bibinfo {year}
  {2011})%
  \bibAnnoteFile{NoStop}{will_coherent_2011}%
\bibitem{trenkwalder_hydrodynamic_2011}%
  \BibitemOpen
  \bibfield{author}{%
  \bibinfo {author} {\bibfnamefont{A.}~\bibnamefont{Trenkwalder}}, \bibinfo
  {author} {\bibfnamefont{C.}~\bibnamefont{Kohstall}}, \bibinfo {author}
  {\bibfnamefont{M.}~\bibnamefont{Zaccanti}}, \bibinfo {author}
  {\bibfnamefont{D.}~\bibnamefont{Naik}}, \bibinfo {author}
  {\bibfnamefont{A.~I.}\ \bibnamefont{Sidorov}}, \bibinfo {author}
  {\bibfnamefont{F.}~\bibnamefont{Schreck}},\ and\ \bibinfo {author}
  {\bibfnamefont{R.}~\bibnamefont{Grimm}},\ }%
  \bibfield{journal}{%
  \Doi{10.1103/PhysRevLett.106.115304}{\bibinfo {journal} {Phys. Rev. Lett.}}\
  }%
  \textbf{\bibinfo {volume} {106}},\ \bibinfo {pages} {115304} (\bibinfo {year}
  {2011})%
  \bibAnnoteFile{NoStop}{trenkwalder_hydrodynamic_2011}%
\bibitem{weber_three-body_2003}%
  \BibitemOpen
  \bibfield{author}{%
  \bibinfo {author} {\bibfnamefont{T.}~\bibnamefont{Weber}}, \bibinfo {author}
  {\bibfnamefont{J.}~\bibnamefont{Herbig}}, \bibinfo {author}
  {\bibfnamefont{M.}~\bibnamefont{Mark}}, \bibinfo {author}
  {\bibfnamefont{H.}~\bibnamefont{N\"{a}gerl}},\ and\ \bibinfo {author}
  {\bibfnamefont{R.}~\bibnamefont{Grimm}},\ }%
  \bibfield{journal}{%
  \bibinfo {journal} {Phys. Rev. Lett.}\ }%
  \textbf{\bibinfo {volume} {91}},\ \bibinfo {pages} {123201} (\bibinfo {year}
  {2003})%
  \bibAnnoteFile{NoStop}{weber_three-body_2003}%
\bibitem{grimm_private}%
  \BibitemOpen
  \bibfield{author}{%
  \bibinfo {author} {\bibfnamefont{R.}~\bibnamefont{Grimm}},\ }%
  \bibinfo {journal} {private communication}%
  \bibAnnoteFile{NoStop}{grimm_private}%
\bibitem{cho_high_2011}%
  \BibitemOpen
\bibfield{journal}{%
    }%
  \bibfield{author}{%
  \bibinfo {author} {\bibfnamefont{H.}~\bibnamefont{Cho}}, \bibinfo {author}
  {\bibfnamefont{D.}~\bibnamefont{{McCarron}}}, \bibinfo {author}
  {\bibfnamefont{D.~L.}\ \bibnamefont{Jenkin}}, \bibinfo {author}
  {\bibfnamefont{M.~P.}\ \bibnamefont{K\"{o}ppinger}},\ and\ \bibinfo {author}
  {\bibfnamefont{S.~L.}\ \bibnamefont{Cornish}},\ }%
  \bibfield{journal}{%
  \bibinfo {journal} {Euro. Phys. Jour. D}\ }%
  \textbf{\bibinfo {volume} {65}},\ \bibinfo {pages} {125} (\bibinfo {year}
  {2011})%
  \bibAnnoteFile{NoStop}{cho_high_2011}%
\bibitem{schuster_avalanches_2001}%
  \BibitemOpen
  \bibfield{author}{%
  \bibinfo {author} {\bibfnamefont{J.}~\bibnamefont{Schuster}}, \bibinfo
  {author} {\bibfnamefont{A.}~\bibnamefont{Marte}}, \bibinfo {author}
  {\bibfnamefont{S.}~\bibnamefont{Amtage}}, \bibinfo {author}
  {\bibfnamefont{B.}~\bibnamefont{Sang}}, \bibinfo {author}
  {\bibfnamefont{G.}~\bibnamefont{Rempe}},\ and\ \bibinfo {author}
  {\bibfnamefont{H.~C.~W.}\ \bibnamefont{Beijerinck}},\ }%
  \bibfield{journal}{%
  \bibinfo {journal} {Phys. Rev. Lett.}\ }%
  \textbf{\bibinfo {volume} {87}},\ \bibinfo {pages} {170404} (\bibinfo {year}
  {2001})%
  \bibAnnoteFile{NoStop}{schuster_avalanches_2001}%
\bibitem{zaccanti_observation_2009}%
  \BibitemOpen
  \bibfield{author}{%
  \bibinfo {author} {\bibfnamefont{M.}~\bibnamefont{Zaccanti}}, \bibinfo
  {author} {\bibfnamefont{B.}~\bibnamefont{Deissler}}, \bibinfo {author}
  {\bibfnamefont{C.}~\bibnamefont{{D{\textquoteright}Errico}}}, \bibinfo
  {author} {\bibfnamefont{M.}~\bibnamefont{Fattori}}, \bibinfo {author}
  {\bibfnamefont{M.}~\bibnamefont{Jona-Lasinio}}, \bibinfo {author}
  {\bibfnamefont{S.}~\bibnamefont{M\"{u}ller}}, \bibinfo {author}
  {\bibfnamefont{G.}~\bibnamefont{Roati}}, \bibinfo {author}
  {\bibfnamefont{M.}~\bibnamefont{Inguscio}},\ and\ \bibinfo {author}
  {\bibfnamefont{G.}~\bibnamefont{Modugno}},\ }%
  \bibfield{journal}{%
  \bibinfo {journal} {Nature Physics}\ }%
  \textbf{\bibinfo {volume} {5}},\ \bibinfo {pages} {586} (\bibinfo {year}
  {2009})%
  \bibAnnoteFile{NoStop}{zaccanti_observation_2009}%
\bibitem{Anmerkung1}%
  \BibitemOpen
  \bibinfo {journal} {Errors are statistical and include contributions from
  errors of atom number, temperature and decay time}%
  \bibAnnoteFile{NoStop}{Anmerkung1}%
\bibitem{wang_private}%
  \BibitemOpen
\bibfield{journal}{%
    }%
  \bibfield{author}{%
  \bibinfo {author} {\bibfnamefont{Y.}~\bibnamefont{Wang}},\ }%
  \bibinfo {journal} {private communication}%
  \bibAnnoteFile{NoStop}{wang_private}%
\bibitem{leblanc_speciespecific_2007}%
  \BibitemOpen
\bibfield{journal}{%
    }%
  \bibfield{author}{%
  \bibinfo {author} {\bibfnamefont{L.~J.}\ \bibnamefont{{LeBlanc}}}\ and\
  \bibinfo {author} {\bibfnamefont{J.~H.}\ \bibnamefont{Thywissen}},\ }%
  \bibfield{journal}{%
  \bibinfo {journal} {Phys. Rev. A}\ }%
  \textbf{\bibinfo {volume} {75}},\ \bibinfo {pages} {053612} (\bibinfo {year}
  {2007})%
  \bibAnnoteFile{NoStop}{leblanc_speciespecific_2007}%
\bibitem{karski_nearest-neighbor_2009}%
  \BibitemOpen
  \bibfield{author}{%
  \bibinfo {author} {\bibfnamefont{M.}~\bibnamefont{Karski}}, \bibinfo {author}
  {\bibfnamefont{L.}~\bibnamefont{F\"{o}rster}}, \bibinfo {author}
  {\bibfnamefont{J.~M.}\ \bibnamefont{Choi}}, \bibinfo {author}
  {\bibfnamefont{W.}~\bibnamefont{Alt}}, \bibinfo {author}
  {\bibfnamefont{A.}~\bibnamefont{Widera}},\ and\ \bibinfo {author}
  {\bibfnamefont{D.}~\bibnamefont{Meschede}},\ }%
  \bibfield{journal}{%
  \Doi{10.1103/PhysRevLett.102.053001}{\bibinfo {journal} {Phys. Rev. Lett.}}\
  }%
  \textbf{\bibinfo {volume} {102}},\ \bibinfo {pages} {053001} (\bibinfo {year}
  {2009})%
  \bibAnnoteFile{NoStop}{karski_nearest-neighbor_2009}%
\end{thebibliography}
%
\end{document}